


\message{ Assuming 8.5" x 11" paper }    

\magnification=\magstep1	          

\raggedbottom

\parskip=9pt

%

\def\singlespace{\baselineskip=12pt}      
\def\sesquispace{\baselineskip=18pt}      


 





 at10pt
 %








\def\tr{\mathop {{\rm \, Tr}} \nolimits}	 



\def\sqr#1#2{\vcenter{
  \hrule height.#2pt 
  \hbox{\vrule width.#2pt height#1pt 
        \kern#1pt 
        \vrule width.#2pt}
  \hrule height.#2pt}}


\def\lto{\mathop
        {\hbox{${\lower3.8pt\hbox{$<$}}\atop{\raise0.2pt\hbox{$\sim$}}$}}}
\def\gto{\mathop
        {\hbox{${\lower3.8pt\hbox{$>$}}\atop{\raise0.2pt\hbox{$\sim$}}$}}}
%


\def\half{{1 \over 2}}


\def\part{\subseteq}		



\def\to{\mathop\rightarrow}	




\def\interior #1 {  \buildrel\circ\over  #1}     




\def\basisvector#1#2#3{
 \lower6pt\hbox{
  ${\buildrel{\displaystyle #1}\over{\scriptscriptstyle(#2)}}$}^#3}

\def\alfa{\alpha}




\fontdimen16\textfont2=2.5pt
\fontdimen17\textfont2=2.5pt
\fontdimen14\textfont2=4.5pt
\fontdimen13\textfont2=4.5pt 




%



\let\miguu=\footnote
\def\footnote#1#2{{$\,$\parindent=9pt\baselineskip=13pt%
\miguu{#1}{#2\vskip -7truept}}}
 %

\def\linebreak{\hfil\break}

\def\pagebreak{\vfil\break}


\def\BulletItem #1 {\item{$\bullet$}{#1 }}
\def\bulletitem #1 {\BulletItem{#1}}

\def\author#1 {\medskip\centerline{\it #1}\bigskip}

\def\address#1{\centerline{\it #1}\smallskip}

\def\section #1 {\bigskip\noindent{\headingfont #1 }\par\nobreak\smallskip\noindent}

\def\subsection #1 {\medskip\noindent{\subheadfont #1 }\par\nobreak\smallskip\noindent}
 %

\def\ReferencesBegin
{
 \singlespace					   
 \vskip 0.5truein
 \centerline           {\bf References}
 \par\nobreak
 \medskip
 \noindent
 \parindent=2pt
 \parskip=6pt			
 }

\def\reference{\hangindent=1pc\hangafter=1} 

\def\ref{\reference}

 %

\def\journaldata#1#2#3#4{{\it #1\/}\phantom{--}{\bf #2$\,$:} $\!$#3 (#4)}
 %

 %


\def\webhome{{\tt http://www.pitp.ca/personal/rsorkin/}}
 %

 %

\def\webtilde{\lower2pt\hbox{${\widetilde{\phantom{m}}}$}}

\def\hpf#1{\webhome{\tt{some.papers/}}}
\def\hpfll#1{\webhome{\tt{lisp.library/}}}
 %

\font\titlefont=cmb10 scaled\magstep2 

\font\headingfont=cmb10 at 12pt
%

\font\subheadfont=cmssi10 scaled\magstep1 






 %



\def\o{{\phantom a}}		

\def\l{\ell}

\def\ket#1{|#1\rangle}      %
\def\bra#1{\langle#1|}      %
\def\tr{\mathop {{\rm \, tr}} \nolimits}	 



\phantom{}


\sesquispace
\centerline{{\titlefont On the Entropy of the Vacuum outside a Horizon}\footnote{$^{^{\displaystyle\star}}$}%
{Except for the present footnote, this is a verbatim copy of an article
 originally published in: 
  B. Bertotti, F. de Felice and A. Pascolini (eds.),
  {\it Tenth International Conference on General Relativity and Gravitation (held Padova, 4-9 July, 1983), Contributed Papers}, 
  vol. II, pp. 734-736 (Roma, Consiglio Nazionale Delle Ricerche, 1983).  
I am grateful to Ted Jacobson for suggesting that I make this article available online.
See also: Luca Bombelli, Rabinder K.~Koul, Joohan Lee and Rafael D.~Sorkin, 
  ``A Quantum Source of Entropy for Black Holes'', 
   \journaldata{Phys. Rev.~D}{34}{373-383}{1986} 
}}

\bigskip


\singlespace			        

\author{Rafael D. Sorkin}
 \address {Center for Theoretical Physics}
 \address{University of Maryland, College Park, MD 20742, U.S.A.}



\bigskip
\sesquispace

The evidence is very strong that a black hole presents itself to the
outside world as a thermodynamic system with entropy proportional to its
surface (horizon) area.  Yet the physical origin of this entropy is far
from clear.  In fact the formula $S=k\,\lg N$, on which our general
understanding of the Second Law is based, entails the absurdity
$S=\infty$; for---unlike in flat space---a bound on the total energy
does not suffice to bound the number of possible internal states.  In
particular the Oppenheimer-Snyder solutions [1] already provide an
infinite number of possible internal configurations for a Schwarzschild
exterior of fixed mass.

A related observation is that the internal dynamics of a black hole
ought to be irrelevant to its exhibited entropy because---almost by
definition---the exterior is an {\it autonomous system} for whose
behavior one should be able to account without ever referring to
internal black hole degrees of freedom.  In particular one should be
able to explain why it happens that a sum of two terms, one referring
to exterior matter and the other only to the black hole geometry, tends
always to increase. 

Based on the conception of the exterior region as an autonomous quantum
system with state given by the density matrix, $\rho=\rho^{ext}$, one
can automatically define an ``exterior entropy''
$S^{ext}=-\tr\rho\lg\rho$.  This paper will estimate a particular
contribution to $S^{ext}$ and show that it does in fact produce a term
proportional to the horizon area.  The other half of the
problem---showing that $S^{ext}$ increases---will not be addressed
(except to point out herewith that on general grounds it suffices to
show that the ``totally random'' state $\rho=1$ evolves into itself. [2])

The contribution we will consider pertains to the quantum fields assumed
to exist in spacetime (including gravitons of course) but will be
estimated only for a non-interacting scalar field.  Specifically
consider a Klein-Gordon field $\phi$ and a Cauchy hypersurface $H^{tot}$
divided into two regions $H^{int}$ and $H^{ext}$, and let the quantum
(mixed) state of $\phi$ with respect to $H^{tot}$ be $\rho^{tot}$.
Tracing out the variables referring to $H^{int}$ produces a reduced
operator $\rho=\rho^{ext}$ which is the effective state for observers
confined to $H^{ext}$.  The entropy 
$S=S^{ext}=-\tr\rho\lg\rho$ is then defined and is in general non-zero
even when $\rho^{tot}$ itself is a pure state (in which case $S$ is an
intrinsically quantum entropy.)  Moreover $S$ depends not at all on the
analogously reduced operator $\rho^{int}$ describing the state of $\phi$
relative to the {\it interior} region $H^{int}$.  Let us estimate $S$ in
the situation that $H^{tot}$ is a $t=$constant hypersurface in Minkowski
space and $\rho^{tot}=\ket{0}\bra{0}$ is the (Minkowski) $\phi$-vacuum.
(This sounds very unlike the situation of physical interest, but in fact
turns out to yield a decent approximation to the latter.)

To begin with we can replace $\phi$ by a lattice of harmonic oscillators
distributed in $H^{tot}$ with density $\l^{-3}$ and whose hamiltonian is a
discretized version of 
$\int\half[(\partial\phi)^2+m^2\phi^2]d^3x$.  With the lattice points
labelled by an index $A$ and $\phi^A$ the value of $\phi$ at the $A^{th}$
such point, the Hamiltonian takes the form
$$
      {1\over 2} G^{AB}p_A p_B + {1\over 2} V_{AB}\phi^A \phi^B
$$
where $G$ and $V$ are both positive definite matrices depending on the
choice of lattice, and in the Schr{\"o}dinger representation 
$p_A = -i\partial/\partial\phi^A$.  Then with the vacuum $\ket{0}$
defined as the minimum energy state, a calculation whose details will
appear elsewhere expresses $S=S^{ext}$ as a sum over the eigenvalues of
a certain operator $\Lambda$: 
$S=\sum_\lambda S(\lambda)$, where, with $\mu$ abbreviating
$1+2\lambda^{-1} - 2[\lambda^{-1} (1-\lambda^{-1})]^{1/2}$,
$S(\lambda) := -\lg(1-\mu) - [\mu/(1-\mu)]\lg\mu$.
$\Lambda$ itself depends on the division of the lattice points into
interior (labelled by $\alfa,\beta$) and exterior (labelled by $a,b$)
and can be expressed as
$\Lambda^a_{\o b} = - W^{a\alfa}W_{\alfa b}$ where $W^{AB}$ is the
inverse matrix of $W_{AB}$, which in turn is the positive square root of
$V_{AB}$ with respect to the scalar product $G^{AB}$: \ 
$W_{AB}G^{BC}W_{CD}=V_{AD}$.
(In the continuum limit $W_{AB}$ is the integral operator whose kernel
is the so-called ``finite part'' of ~$-\pi^{-2}|x-y|^{-4}$.)

On dimensional grounds it is easy to see that $S$ will be ultra-violet
infinite in the continuum limit $\l\to 0$.  For finite $\l$ the fact
that $S$ is a {\it sum} over the eigenvalues of $\Lambda$, coupled with
the fact that the singularity in $\Lambda=\Lambda(x,y)$ occurs at $x=y$,
means that the leading term in $S$ will be proportional to $A/\l^2$
where $A$ is the area of the surface (``horizon'') separating $H^{int}$
from $H^{ext}$.  The proportionality constant can be estimated as
$3\int_0^1 y \, dy \, \sigma(y^2+m^2\l^2)$ where $\sigma(y^2)$ is the value of
$S^{ext}$ computed in $1+1-$dimensions with the mass equal to $y$ in
units of the inverse lattice spacing and with $H^{ext}$ taken to be a
half-line.  (Since for $y\to 0$ we have $-y\sigma(y^2)\sim y\lg y\to0$,
$S$ is to leading order independent of $m$, as one would expect.)

To obtain an entropy of the correct order of magnitude for a black hole,
the cutoff $\l$ must be chosen approximately equal to the Planck length.
Conversely, if $S^{ext}$ really can be identified as the black hole
entropy we obtain evidence of the physical necessity for such a cutoff
to exist.

However caution is indicated by several circumstances in addition to the
obvious one that it remains to be explained why $S^{ext}$ necessarily
increases with time.  Aside from this, the most serious problem seems to
be that for free fields $S^{ext}$ would be proportional to the number of
massless fields, $\l$ being fixed.  If this dependence is not cured by
the presence of large couplings at high energies (and it may well not be
in asymptotically free theories) or by a (supersymmetric?) conspiracy
relating $\l$ to the number and type of fundamental fields, then it must
be cured by taking into account the coupling between the fields and the
horizon shape itself (``back reaction'').  This in any case must
introduce an $\l$ limiting the validity of the above semi-classical
computation, and moreover limiting it in such a way as to tend to cancel
the unwanted dependence on the number of fields. 
Beyond evaluating this cutoff (which should be done) one could try to
take the next logical step of evaluating the degrees of freedom of the
horizon itself, which one might argue also are about
$\exp(A/\l^2_{Planck})$ in number.  For if the entropy can't be inside
the black hole and proves not to be outside it either, then where else
can it possibly be but the horizon?

\ReferencesBegin                             

\ref [1] Misner, C.W., K.S. Thorne and J.A. Wheeler (1970) 
{\it Gravitation}, San Francisco: W.H. Freeman

\ref [2] Schnakenberg, J. (1976) Rev. Mod. Phys. {\bf 48} 571

\end 

(prog1 'now-outlining
  (Outline* 
     "\f"                   1
      "
      "
      "
      "\\Abstract"          1
      "\\section"           1
      "\\subsection"        2
      "\\ReferencesBegin"   1
      "
      "\\ref "              2
      "\\end